\documentstyle[preprint,prl,aps]{revtex}

\tightenlines

\input epsf

\title{Phase Transition in Hot Pion Matter}

\author{
A.~Kostyuk$^{1,2}$, M.~Gorenstein$^{1,2}$,
H.~St\"ocker$^1$  and W.~Greiner$^1$
}

\address{
$^1$ Institut f\"ur Theoretische Physik, Universit\"at  Frankfurt,
Germany\\
$^2$ Bogolyubov Institute for Theoretical Physics,
Kyiv, Ukraine
}

\begin{document}

\maketitle

\begin{abstract}
The equation of state for the pion gas is analyzed within the third
virial approximation. The second virial coefficient is found 
from the $\pi \pi$-scattering data, while the third one is
considered as a free parameter. 
The proposed model leads to  
a first-order phase transition from the pion gas to a more dense 
phase at the temperature $T_{pt}<136$ MeV. Due to relatively low
temperature this phase transition cannot be related to the  
deconfinement. This suggests that a new phase of hadron
matter --- 'hot pion liquid' --- may exist. 
\end{abstract}


\vskip 0.5cm

The widely used thermal models of the hadron gas (HG) are mostly based 
on the ideal gas approach. To take into account attractive interactions 
between hadrons, the hadron resonances are included in the ideal
gas picture \cite{Dashen69}. 
This method, describing quite satisfactory  the gaseous
phase, does not lead  to a phase transition (PT).
In phenomenological models, PT is usually introduced `by hand':
it is postulated from the very beginning that two different phases,
for instance HG and quark gluon plasma, exist; each phase has its own
equation of state and Gibbs criterion is used to find the PT 
point.

It is interesting to study whether a PT can naturally follow 
from the equation of state of real HG
without any {\it ad hoc} assumptions. This would allow us to describe, 
at least at a qualitative level, both phases within a unique
approach and to estimate the PT parameters using available hadron data.  

In the present letter we make the first step in this direction:
we study the equation of state of the {\it pion} matter, i.e. we ignore 
(anti-)nucleons and strange particles.

Our starting point is
the equation of state for a gas of interacting particles 
in the form of virial expansion \cite{Huang,Mayer,Greiner}:
\begin{equation} \label{virexpanP}
p(T, n) = T \sum_{i=1}^{\infty} a_i(T) n^i.
\end{equation}
It allows to calculate the gas pressure $p$ at given temperature
$T$ and particle density $n$, provided that the virial coefficients
$a_i(T)$ are known. 

In the case of a relativistic system, like a hot pion gas, 
the particle number is not conserved, hence it is inconvenient to
choose $n$ as an independent variable. On the other hand,  
the chemical potential  $\mu$ for pions in isotopically symmetric systems 
at chemical  equilibrium is equal to zero. Therefore, 
it is more convenient to choose $T$ and $\mu$ as independent variables:
$n=n(T,\mu=0)$ should be found from Eq.(\ref{virexpanP}) in thermodynamically
consistent way.

Let us express the thermodynamical quantity of the gas satisfying
Eq.(\ref{virexpanP}) via $T$ and $\mu$. 
The particle density is given by the formula
\begin{equation} \label{dens}
n(T,\mu) = \frac{\partial p(T, \mu)}{\partial \mu}.
\end{equation}
Using the equation of state (\ref{virexpanP}) leads to a differential
equation for $n(T,\mu)$: 
\begin{equation} \label{diffeq}
n(T,\mu) = 
\frac{\partial p(T, n)}{\partial n}  
\frac{\partial n(T, \mu)}{\partial \mu} =
T \sum_{i=1}^{\infty} i a_i(T) n^{i-1}(T, \mu) 
\frac{\partial n(T, \mu)}{\partial \mu}.
\end{equation}
Integration of this equation yields
\begin{equation} \label{sol}
\mu =
T \left[ a_1(T) \log \left[ C(T) n(T, \mu) \right] +
\sum_{i=2}^{\infty} \frac{i}{i-1} a_i(T) n^{i-1}(T, \mu) 
\right],
\end{equation}
where $C(T)$ is the integration constant. In the low density limit
$n \rightarrow 0$ at fixed temperature (which is equivalent to 
$\mu \rightarrow - \infty$) the equation (\ref{sol}) should reproduce the 
properties of the ideal classical gas, i.e. $n(T, \mu)$ should
coincide with the absolute activity $z(T,\mu)$, which is given
by the formula \cite{Mayer}:
\begin{equation} \label{z}
 z(T,\mu) \equiv g \phi(T;m)  \exp\left( \frac{\mu}{T}  \right),
\end{equation}
where
$g$ is the (iso-)spin degeneracy factor ($g=3$ for the pion),
$m$ is the pion mass,
and $\phi(T;m)$ can be expressed via the $K_{2}$ modified Bessel function 
\begin{equation}\label{phi}
\phi(T;m)~ =~ \frac{1}{2 \pi^2} \int_0^{\infty}p^2 dp~
\exp \left( - \frac{\sqrt{p^{2}+m^{2}}}{T}  \right)~
= ~\frac{m^{2} T}{2 \pi^{2}}~ K_{2}\left( \frac{m}{T} \right)~.
\end{equation}
This allows to fix the integration constant. Taking into account that 
$a_1(T) \equiv 1$ one gets
\begin{equation} \label{C}
 C(T) = \frac{\exp(\frac{\mu}{T})}{z(T,\mu)} \equiv  
 \frac{1}{g \phi(T;m)} \ ,
\end{equation}

Finally, one has the following transcendental equation for 
$n(T,\mu)$
\begin{equation} \label{transc}
 z(T,\mu) = n(T,\mu) \exp \left[
 \sum_{i=1}^{\infty} \frac{i+1}{i} a_{i+1}(T) n^i(T,\mu)
 \right].
\end{equation}

\begin{figure}[t]
\begin{center}
\vfill
\leavevmode
\epsfysize=20cm \epsfbox{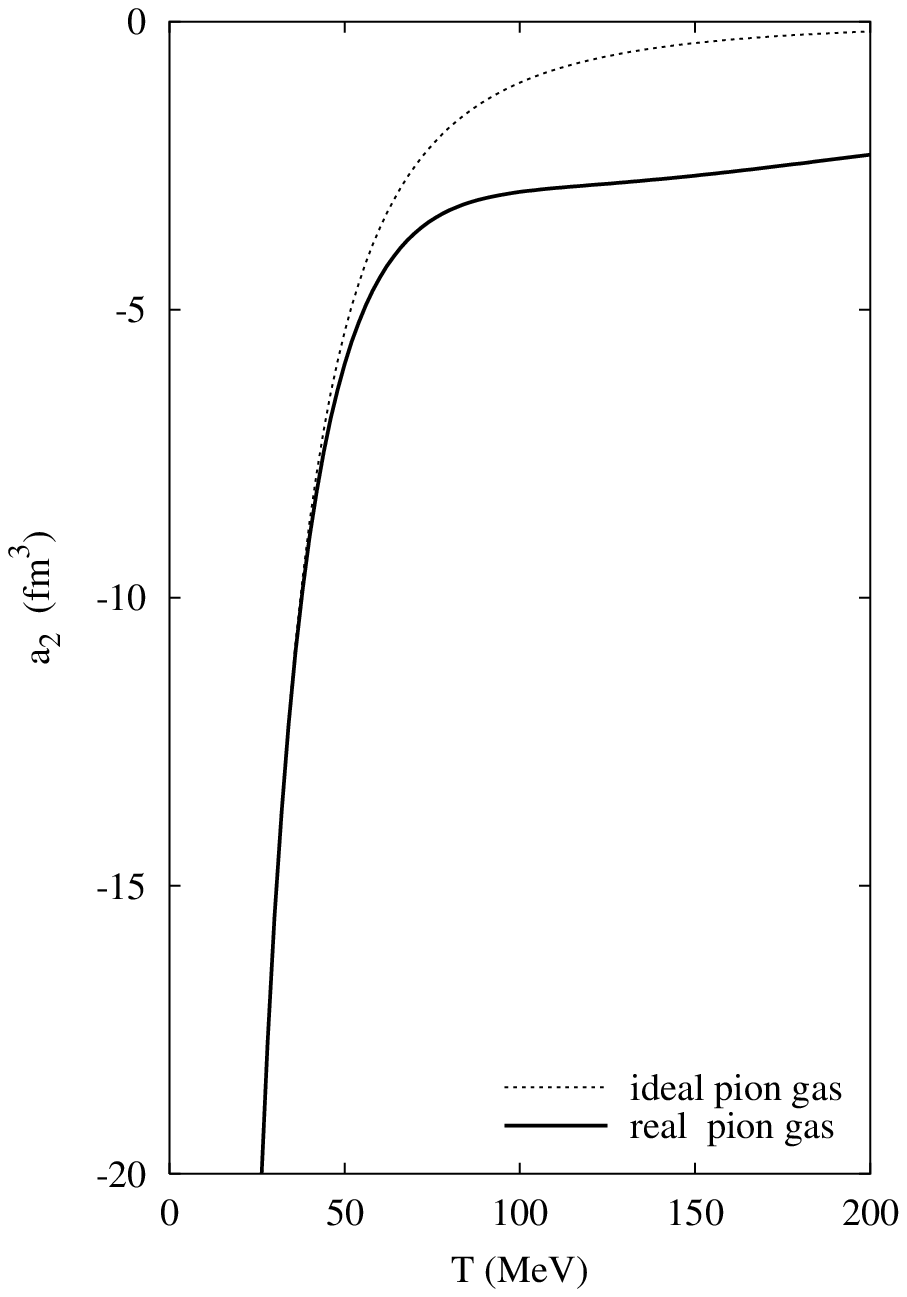}
\vfill
\mbox{}\\
\caption{The dependence of the second virial coefficient of the 
ideal and real pion gas on the temperature.
\label{fa2}
}
\end{center}
\end{figure}

To apply this equation to the pion gas it is necessary to know the 
values of the virial coefficients. The second virial coefficient is straightly
related to the second Mayer's cluster integral \cite{Huang,Mayer,Greiner}: 
$a_2=-b_2$, which is given by 
\begin{equation} \label{b2}  
  b_2 = b_2^{(0)} + b_2^{(i)},
\end{equation}
where the first term 
\begin{equation} \label{b20}  
  b_2^{(0)} = \frac{1}{2 g} \  
  \frac{\pi^2}{2 g m^2 T} \ \frac{K_2(2 m/T)}{[ K_2(m/T) ]^2}
\end{equation}
appears due to Bose effects. It exists in an ideal Bose   
gas too. The second term  
\begin{equation} \label{b2i}  
  b_2^{(i)} = \frac{2 \pi}{ m^4 T^2 [ g K_2(m/T) ]^2}
  \sum_{I=0}^{2}  
 {\sum_{L}}^\prime (2I+1) (2L+1)
 \int_0^\infty d \varepsilon  \varepsilon^2
 \delta_{L,I}(\varepsilon)   
 K_1 \left( \frac{\varepsilon}{T} \right).
\end{equation}
is related to the interactions
between particles. The function $\delta_{L,I}(\varepsilon)$ stands 
for the phase shift \cite{Hyams73,Ishida96,Ishida97,Ishida99} of 
$\pi \pi$-scattering in the state with angular momentum
$L$, isospin $I$ and center-of-mass energy $\varepsilon$.
The sum over $L$ runs over those values, which ensure 
symmetry of the total two pion wave function, i.e.
\begin{equation} \label{Lval} 
 L = \left\{
   \begin{array}{lll}
   0,2,4,6, \dots & \mbox{for} & I=0,2  \\
   1,3,5,7, \dots & \mbox{for} & I=1 .  \\ 
   \end{array} 
 \right.
.
\end{equation}
The second cluster integral $b_2$ was calculated in \cite{Kostyuk:2000nx}.
The dependence of $a_2$ on the temperature is shown in Fig.\ref{fa2}
(see Ref.\cite{Kostyuk:2000nx} for details).

A model independent determination of all other virial 
coefficients $a_i(T)$, $i > 2$ is hardly possible. The quantum theory is 
developed only for $a_3(T)$ \cite{Pais59}, but lacking of detailed 
experimental information on the three-pion interaction makes it impossible 
to use this theory for the case of pion gas. For the higher virial 
coefficients, the problem looks even more hopeless.  

A simple model will be adopted for the further consideration:
\begin{itemize}
\item  The virial series is truncated after the third term: 
       $a_i(T) \equiv 0$, $i > 3$;
\item  The third virial coefficient $a_3(T)$, at fixed $T$,
       is considered as a free parameter. 
\end{itemize}

It is convenient to introduce the notation 
$\xi(T) \equiv a_3(T)/[ a_2(T) ]^2$ and treat the dimensionless quantity
$\xi(T)$ as a free parameter. 
In this case, the equation of state (\ref{transc}) takes the form 
\begin{equation} \label{modp}
p(T, n(T,\mu)) = T n(T,\mu) 
\left \{
1 +  a_2(T) n(T,\mu) + \xi(T) [ a_2(T) n(T,\mu) ]^2 \right \},  
\end{equation}   
where the value of $n(T,\mu)$ is found from the transcendental 
equation
\begin{eqnarray} \label{modn}
z(T,\mu) &=& n(T,\mu) \exp \left[ 2 a_2(T) n(T,\mu) + 
\frac{3}{2} \xi(T) \left[ a_2(T) n(T,\mu) \right]^2
 \right] .
\end{eqnarray}

\begin{figure}[t]
\begin{center}
\vfill
\leavevmode
\epsfysize=20cm \epsfbox{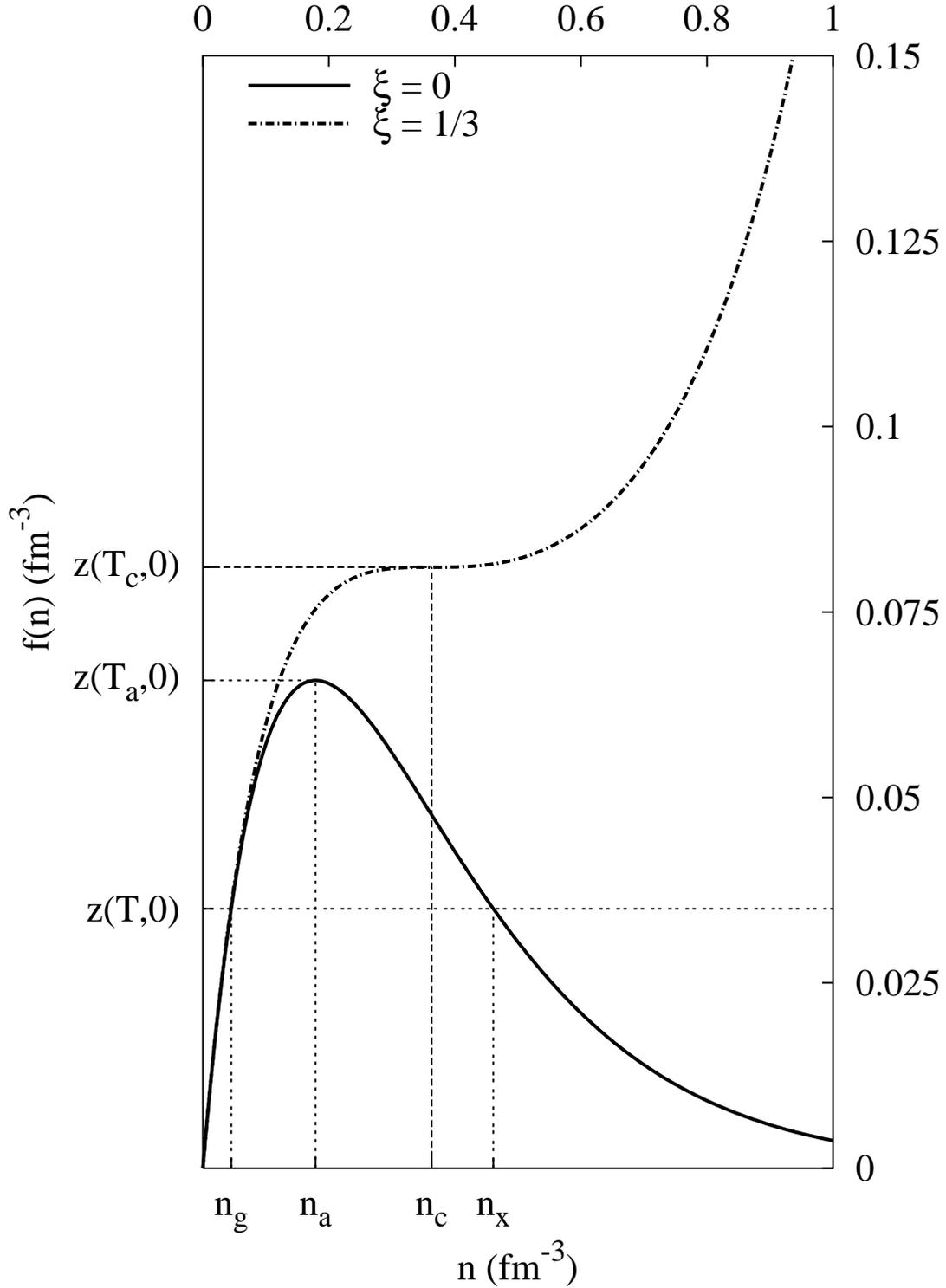}
\vfill
\mbox{}\\
\caption{The dependence of the r.h.s. of the equation  (\ref{modn})
$f(n)$ on the variable $n$ at $\mu=0$ for $\xi(T) = 0$ and $\xi(T) = 1/3$ 
at $T = T_a = 128$ MeV and $T=T_c=136$ MeV, 
respectively.
The behavior of $f(n)$ for $\xi(T) < 0$ ($\xi(T) > 1/3$) is qualitatively
the same as for $\xi(T) = 0$ ($\xi(T) = 1/3$). 
\label{fnopt}
}
\end{center}
\end{figure}

\begin{figure}[t]
\begin{center}
\vfill
\leavevmode
\epsfysize=20cm \epsfbox{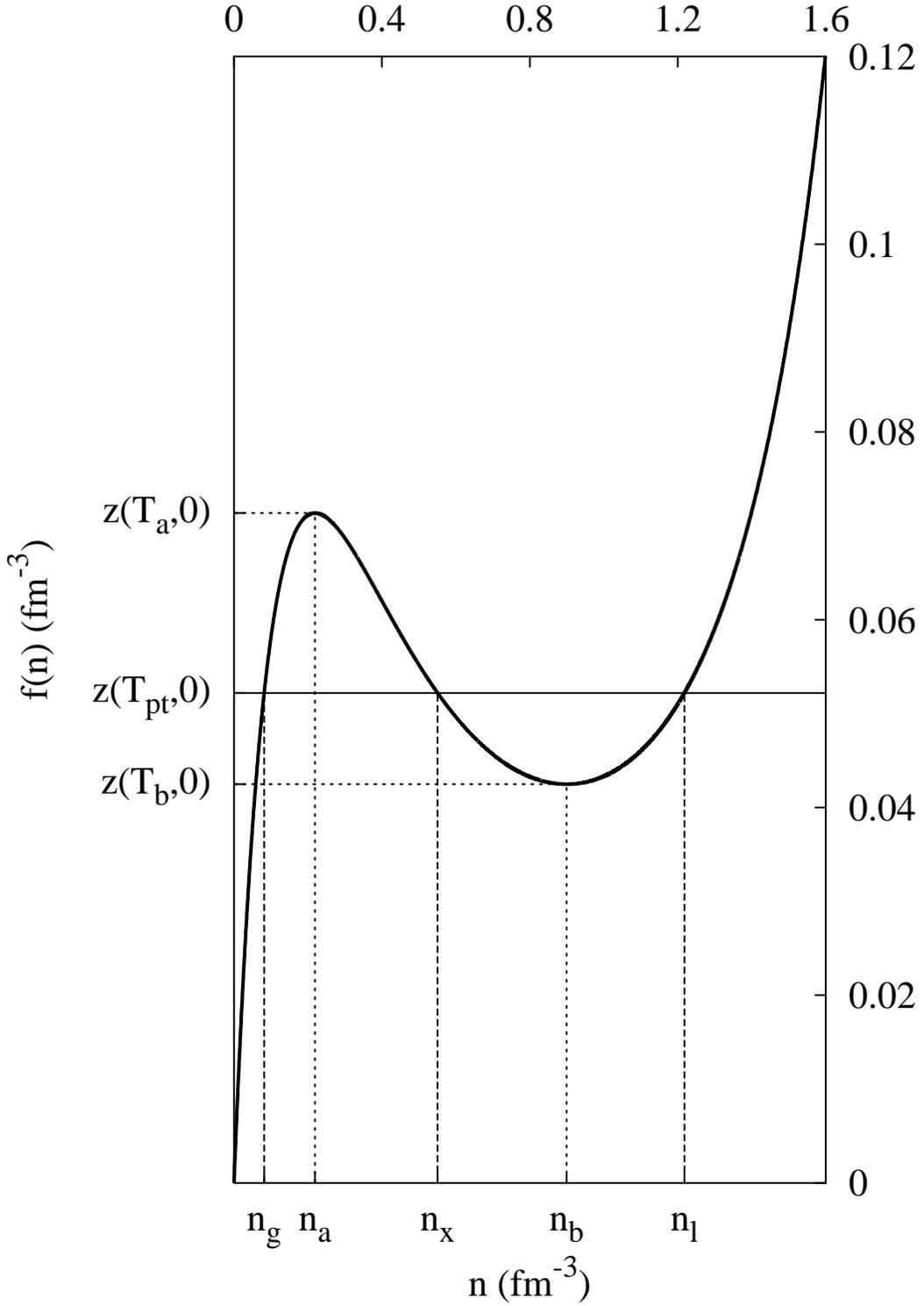}
\vfill
\mbox{}\\
\caption{The dependence of the r.h.s. of the equation  (\ref{modn})
$f(n)$ on the variable $n$ at $\mu=0$ for $\xi(T) = 0.21$ at the
temperature $T=T_{pt}=120$ MeV.
The behavior of $f(n)$ for any $\xi(T)$ from the interval 
$0 < \xi(T) < 1/3$ at any temperature is qualitatively the same. 
\label{fpt}
}
\end{center}
\end{figure}

Let us analyze the behavior of the r.h.s. of Eq.(\ref{modn}) at some
temperature $T$, as a function of $n>0$. For convenience we denote 
it by $f(n)$. Three distinct cases are possible (see Figs. \ref{fnopt} 
and \ref{fpt}):
\begin{enumerate}
\item $\xi(T) \le 0$. The function $f(n)$ has a single extremum --- 
a maximum at the point
\begin{equation}
n_a = \left\{
\begin{array}{lll}
\frac{1}{2 |a_2|} & \mbox{ if } & \xi(T)=0 \\ 
\mbox{} \\
\frac{\sqrt{1+3 |\xi(T)|}-1}{3 |\xi(T)| |a_2|}
& \mbox{ if } & \xi(T) < 0,
\end{array}
\right.
\end{equation}
\item  $\xi(T) \ge 1/3$. The
function $f(n)$ grows monotonically for all $n>0$.
\item  $0 < \xi(T) < 1/3$. In this case the function $f(n)$ has two
extrema: a maximum at the point 
\begin{equation}
n_a = \frac{1-\sqrt{1 - 3 \xi(T)}}{3 \xi(T) |a_2|}
\end{equation} 
and a minimum at the point
\begin{equation}
n_b = \frac{1+\sqrt{1 - 3 \xi(T)}}{3 \xi(T) |a_2|}.
\end{equation}  
\end{enumerate} 

In the {\bf first} case, the values of $f(n)$ are bound from the 
above: $f(n) \le f(n_a)$.   
As $z(T,\mu)$ is a monotonically increasing function of the
temperature, the equation 
(\ref{modn}) has no solution at high temperatures $T > T_a$,
where the temperature $T_a$ is found from the equation
\begin{equation}\label{Ta}
z(T_a,\mu)=f(n_a). 
\end{equation}
The numerical calculations give 
$T_a \le 128$ MeV assuming chemical equilibrium ($\mu=0$). 
(The equality is reached at $\xi(T_a)=0$.) 
This means that if 
the third virial coefficient of the pion gas
is negative at the temperatures $T \agt 128$ MeV, 
the model cannot describe the pion matter at these temperatures. 
Consideration of higher order terms of virial expansion would be necessary 
to obtain a satisfactory equation of state of the pion gas in this case.
  
At $T < T_a$ the equation (\ref{modn}) has two solutions: $n_g$ and $n_x$,
$n_g < n_a < n_x$ (See Fig. \ref{fnopt}). It is easy to check that the 
derivative of the pressure with respect to the density is negative 
at the point $n_x$:
\begin{equation}
\left. \frac{\partial p(T,n)}{\partial n} \right|_{n=n_x} < 0.  
\end{equation}
This means that the root $n_x$ corresponds to an unstable phase,
which cannot exist, because any small fluctuations of the 
density would destroy it. The solution $n_g$ corresponds to a stable phase 
which can be obviously interpreted as a pion gas.

The interpretation of the {\bf second} case is straightforward. 
The equation (\ref{modn}) has a unique solution at any temperature,
which describes the gaseous phase of the pion matter. No PT takes place.

The {\bf third} case is the most interesting. The equation 
(\ref{modn}) has solutions at any temperature. 
At low temperatures, $T < T_b$, where $T_b$ satisfies the condition
$z(T_b,\mu)=f(n_b)$, the equation (\ref{modn}) has a single solution
$n_g$ ($n_g < n_a$), which can be obviously interpreted as the density
of gaseous phase. As it is shown in Fig. \ref{fpt}, 
at intermediate temperatures $T_b < T < T_a$, $T_a$ is given by Eq.(\ref{Ta}),
two more solutions appear: $n_x$ ($n_a < n_x < n_b$) and 
$n_l$ ($n_l > n_b$). Similar to the {\bf second} case, the root $n_x$ 
corresponds to an unstable phase, which cannot exist.
In contrast, the points $n_g$ and $n_l$ give
a positive derivative of the pressure and correspond to  
(meta-)stable phases. The questions which of two phases is stable 
or metastable is resolved by the Gibbs criterion: the pressure of a
stable phase is higher than that of metastable one. The point, when the
two pressures are equal to each other  
\begin{equation}
p(T_{pt}, n_g(T_{pt},\mu)) = p(T_{pt}, n_l(T_{pt},\mu))  
\end{equation}
is a PT point. It is reminiscent of the first order
gas-liquid PT. At high 
temperatures $T > T_a$ the equation (\ref{modn}) has an unique solution
corresponding to this `liquid' phase.  

In spite of its similarity to the gas-liquid PT, which
is well known from molecular physics , there is an essential difference.
In molecular physics, the stable liquid exists at temperatures 
below the PT point ($T < T_{pt}$), while high
temperatures ($T > T_{pt}$) correspond to the gaseous phase.   
Here we observe the opposite situation: at low temperatures ($T < T_{pt}$)
the pion matter only slightly differs from 
the ideal classical pion gas. The dense 'liquid' phase appears at high
($T > T_{pt}$) temperatures and thus can be called `hot pion liquid'.


There are two reasons that explain the qualitative difference in the behavior  
of the pion matter as compared with ordinary substances known from 
molecular physics:  
\begin{itemize}
\item In contrast to the number of molecules, the number of pions is not
conserved and rapidly decreases as temperature falls down. So that the 
density of isotopically symmetrical pion matter becomes
so small  at low temperatures that the interactions between the particles 
plays almost no role, which corresponds to a perfect gas.     
\item  In molecular physics the dependence of the second virial 
coefficient on the temperature typically can be approximated by 
the formula \cite{Greiner} 
$a_2(T) = \beta - \frac{\alpha}{T}$, where nonnegative 
constants $\alpha$ and $\beta$ come, respectively, from attractive and 
repulsive parts of the intermolecular potential. It is easily seen that 
at high temperatures the repulsive part becomes dominating and the 
virial coefficient becomes positive. 
In contrast, the second virial 
coefficient of the pion gas remains negative at high
temperatures (see Fig.\ref{fa2}), which indicates the attractive nature
of $\pi \pi$-interaction.  
\end{itemize}

As was mentioned above, we study the pion matter at chemical equilibrium.
Thus the dense phase of pion matter is distinct from multipion droplets 
considered in Ref.\cite{Mishustin}. They may exist only at $\mu > 0$ at 
low temperature and evaporate if the temperature exceeds some critical 
value, similarly to ordinary liquid.

\begin{figure}[t]
\begin{center}
\vfill
\leavevmode
\epsfysize=20cm \epsfbox{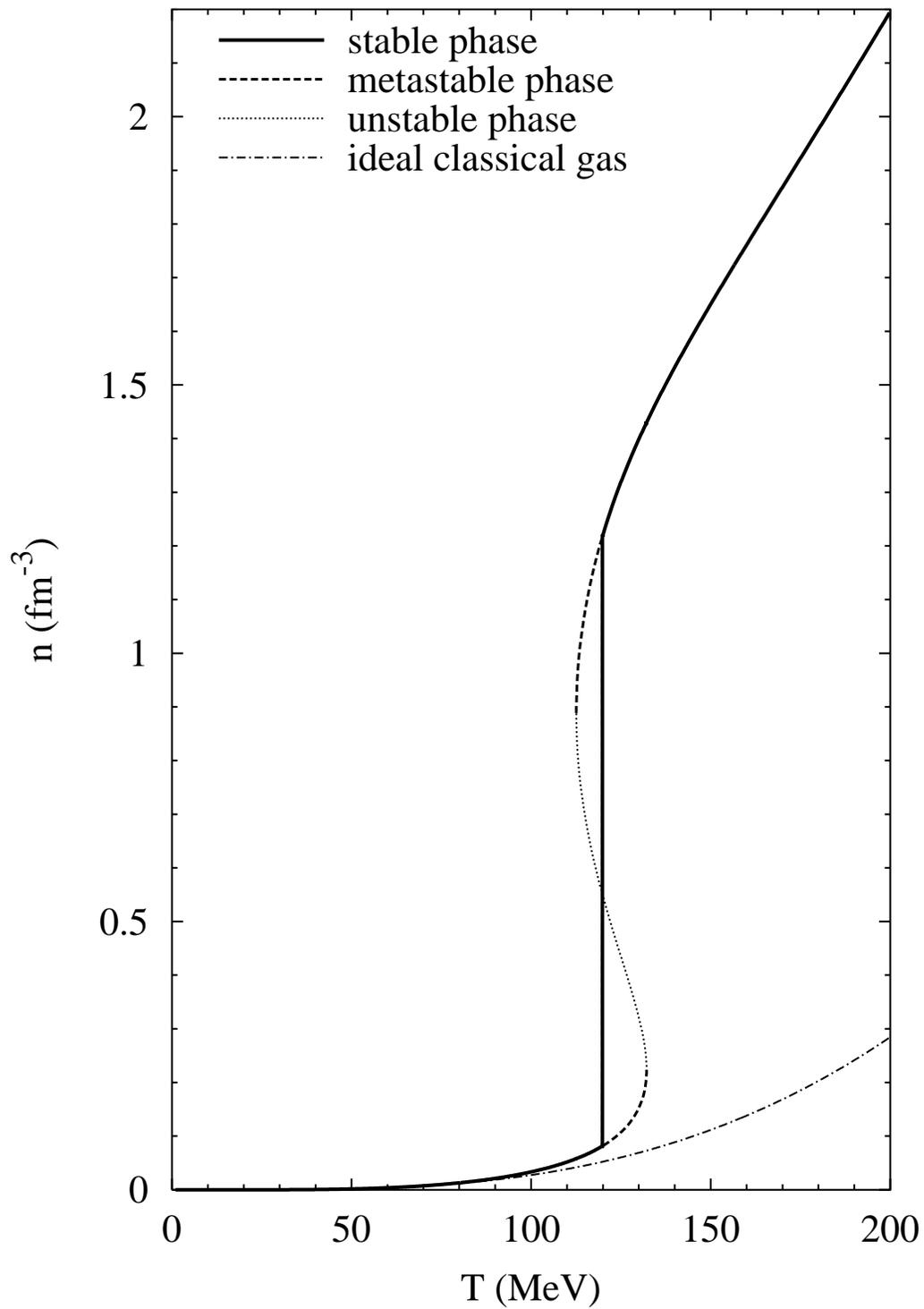}
\vfill
\mbox{}\\
\caption{The dependence of the pion density on the temperature
at chemical equilibrium $\mu=0$ for $\xi(T) = 0.21$. The density of the ideal
classical pion gas is shown for comparison.  
\label{nT}
}
\end{center}
\end{figure}

The dependence of the particle density on the temperature is shown 
in Fig. \ref{nT}. The value of the free parameter $\xi(T) \equiv 0.21$ was 
arbitrary chosen from the interval $0 < \xi(T) < 1/3$ to provide an example.
As is seen from the figure, the density of the gaseous phase is only 
slightly different from the ideal gas density indicating that virial expansion
can be safely applied in this region. On the other hand, the density of
the 'liquid' phase is about one order of magnitude higher than that
of the pion gas, so one should not expect that the present model 
can provide quantitative description of the 'pion liquid'. It rather 
illustrates the features of PT at a qualitative level. 

At the same time, to estimate the upper bound of the PT 
temperature it is sufficient to analyze the properties of the gaseous 
phase. In fact, $T_{pt}$ cannot exceed the maximum temperature 
of the overheated gas $T_{pt} < T_a$ (see Fig. \ref{fpt}).
The upper bound for $T_a$ is approached
at $\xi(T_a) \rightarrow 1/3$ and comprises $136$ MeV. At $\xi(T) \ge 1/3$ the
phase transition, as was mentioned above, disappears. That is, { \bf
our model states that if a first order PT does exist in the 
isotopically symmetric equilibrium pion matter, it takes place at the
temperature not higher than $136$ MeV. } 

A similar first order PT in hadron matter but at higher 
temperature ($T=190 \div 200$ MeV) was predicted long ago within 
various models \cite{Hagedorn:1980cv,Stoecker}. In 
Ref.\cite{Hagedorn:1980cv} the hot and dense phase
was interpreted as quark-gluon plasma, i.e. the PT was 
identified with color deconfinement PT. This agrees with
the estimations of  the temperature of the 
deconfinement PT from lattice simulations of two-flavor QCD:  
$T_{dec} = 150$ -- $200$ MeV
\cite{Blum:1995zf,Greiner:1996wv}.

In contrast, the present model predicts a  PT at  
lower temperature, which cannot be identified  with the 
deconfinement PT. This suggest the possibility that a new phase of 
hadron matter, distinct from the HG 
and QGP may exist at zero baryonic chemical potential in the 
temperature range $136$ MeV $ \alt T \alt T_{dec}$.
It is formed from colorless pions attracting to one another
rather than from deconfined quarks and gluons and can be called 'hot
pion liquid'.

It is possible that the 'hot pion liquid' can be created in relativistic 
heavy ion collisions at RHIC and LHC energies, where the pions will play a 
dominant role at the final stage of the reaction. 

\mbox{}\\

The authors are grateful to Larry McLerran for fruitful discussion and his
interest to the paper and to Gephard Zeeb for reading the manuscript and 
usefull comments.  
The financial support of GSI and DAAD (Germany) is appreciated.
The research described in this publication was made possible in part by
Award No. UP1-2119 of the U.S. Civilian Research \& Development
Foundation for the Independent States of the Former Soviet Union
(CRDF).

\end{document}